\documentclass[prc,aps,manuscript]{revtex4}
\usepackage{graphicx}
\usepackage{dcolumn}
\usepackage{bm}

\newcommand{\bs}{\bigskip}
\newcommand{\bc}{\begin{center}}
\newcommand{\ec}{\end{center}}

\begin{document}

\title{Collective motion in quantum diffusive environment}
\author{ V. M. \surname{Kolomietz}$^{1}$
\footnote{Electronic address: vkolom@kinr.kiev.ua}}
\author{ S. \surname{\AA berg}$^{2}$
\footnote{Electronic address: Sven.Aberg@matfys.lth.se}}
\author{ S. V. \surname{Radionov}$^{1,2}$
\footnote{Electronic address: Sergey.Radionov@matfys.lth.se}}
\affiliation{$^1$ \textit{Institute for Nuclear Research, 03680 Kiev, Ukraine} \\
$^2$ \textit{Department of Mathematical Physics, LTH, SE-221 00 Lund, Sweden}
}
\date{\today}

\begin{abstract}
The general problem of dissipation in macroscopic large-amplitude collective
motion and its relation to energy diffusion of intrinsic degrees of
freedom of a nucleus is studied. By applying the cranking approach to the
nuclear many-body system, a set of coupled dynamical equations for the
collective classical variable and the quantum mechanical occupancies of the
intrinsic nuclear states is derived. Different dynamical regimes of the
intrinsic nuclear motion and its consequences on time properties of
collective dissipation are discussed.
\end{abstract}

\maketitle


\section{Introduction}

The appearance of dissipation for large--amplitude collective motion in
nuclei is still an unsolved problem. The transport models of the nuclear
collective motion like the linear response theory~\cite{lrs1} or the
wall--formula approach~\cite{wfa} assume a priori that the collective dynamics
is adiabatically slow, such that the fast intrinsic nucleonic subsystem
has always sufficient time to adjust to the large changes of collective
deformation parameters. In that case one can say that statistical
equilibrium for the fast intrinsic subsystem is established instantaneously
providing the essentially Markovian equations of motion for the collective
variables.

In the general case the adiabaticity of the collective motion must not
be implied a priori, and one should consider selfconsistently dynamics
of the collective and intrinsic nucleonic degrees of freedom. This is
quite important when we are dealing with nuclear fission at high
excitation energies or the initial stage of heavy ion collisions, i. e.,
when the typical times for the macroscopic collective and intrinsic nucleonic
motions are of the comparable size. Here one would rather expect a
non--Markovian collective dynamics caused by the complex energy flow
between the macroscopic collective and intrinsic nucleonic modes.

Memory (non--Markovian) effects in a time evolution of the collective
parameters have been studied within the linear response theory~\cite{lrs2},
the time--dependent shell--model theory~\cite{tdsmt}, Fermi--liquid
model~\cite{book,kosh04} and etc. If in all these approaches the main
focus is made on the non--Markovian collective motion, we shall
concentrate on the selfconsistent description of the dynamics of the
collective and nucleonic degrees of freedom. The complex
intrinsic nuclear motion at the high excitation energies can be
described within random matrix theory. This provides a measure
how different dynamic regimes of the nucleonic excitations show up
in the corresponding dissipative properties of the collective motion.

The plan of the paper is as follows. In Sect.~\ref{nmbs} we start from the
cranking approach to nuclear many--body problem. Sect.~\ref{iqdd} is devoted
to the quantum--mechanical description of the intrinsic nuclear excitations.
In Sect.~\ref{idcd}, we derive a system of coupled equations for the slow
collective and fast intrinsic modes of the nuclear many--body motion and
measure how the energy diffusion of the quantum--mechanical occupancies of
the nuclear states defines the time properties of the collective friction.
We apply our model to the description of nuclear fission dynamics on the
part of descent from fission barrier to scission point in Sect.~\ref{nfc}.
Finally, conclusions and discussion of the main results of the paper are
given in the Summary.

\section{Nuclear many-body system}

\label{nmbs}

The total energy of the nucleus under collective excitation $\Xi _{tot}$
may be written as
\begin{equation}
\Xi _{tot}=E_{pot}(q)+\frac{1}{2}B(q)\dot{q}^{2}+E^{\ast }(t),  \label{Etot}
\end{equation}%
where $q(t)$ is a single classical collective variable (a "nuclear
deformation"), $E_{pot}$ is the collective potential energy,
$B$ is the collective mass coefficient
and $E^{\ast }$ is the excitation energy of the intrinsic nucleonic degrees
of freedom. Writing the energy of the nucleus in the form of Eq.~(\ref{Etot}%
), we pick out explicitly the contribution from the virtual
transition between the nuclear states, which gives rise to a
collective kinetic energy term $(1/2)B\dot{q}^{2}$, and the
contribution from the real nuclear transitions leading to the
intrinsic excitation energy $E^{\ast }$.

Since the total energy of the nucleus is conserved, we can
derive an equation of motion for the classical collective variable $q$ by
differentiating with respect to time the both sides of Eq.~(\ref{Etot}),
\begin{equation}
B(q)\ddot{q}=-\frac{1}{2}\frac{\partial B(q)}{\partial q}\dot{q}^{2}-\frac{%
\partial E_{pot}(q)}{\partial q}-\frac{1}{\dot{q}}\frac{dE^{\ast }(t)}{dt}.
\label{colem}
\end{equation}%
To study how the dissipation in the collective motion may arise, we shall
derive an expression for the intrinsic excitation energy $E^{\ast }$.

\section{Intrinsic quantum diffusive dynamics}

\label{iqdd}

We treat intrinsic nucleonic motion of the nucleus quantum--mechanically and
start from the Liouville equation for the density matrix operator $\hat{\rho}
$,
\begin{equation}
i\hbar \frac{\partial \hat{\rho}}{\partial t}=\left[ \hat{H}(q[t]),\hat{\rho}%
\right] ,  \label{rho}
\end{equation}%
where $\hat{H}$ is the nuclear many--body Hamiltonian. A moving basis is
introduced as eigenstates of the time--independent Hamiltonian,
\begin{equation}
\hat{H}(q)\Psi _{n}(q)=E_{n}(q)\Psi _{n}(q).  \label{movbas}
\end{equation}%
That is determined by a set of the static many--body wave functions $\Psi _{n}$
and energies $E_{n}$ found at each fixed value of the collective variable $q$.

Now one can rewrite Eq.~(\ref{rho}) as
\begin{equation}
i\hbar \frac{\partial \rho _{nm}}{\partial t}=\sum_{n}\left\{ W_{lm}\rho
_{nl}-W_{nl}\rho _{lm}\right\}  \label{rnm}
\end{equation}%
with
\begin{equation}
\rho _{nl}=\exp \left( -i\int_{0}^{t}\omega _{nl}(q[t^{\prime }])dt^{\prime
}\right) \langle \Psi _{n}|\hat{\rho}|\Psi _{l}\rangle ,  \label{rnl}
\end{equation}%
\begin{equation}
W_{nl}=\exp \left( -i\int_{0}^{t}\omega _{nl}(q[t^{\prime
}])dt^{\prime }\right) \langle \Psi _{n}|i\frac{\partial
}{\partial t}|\Psi _{l}\rangle , \label{wnl}
\end{equation}%
and $\omega _{nl}=(E_{n}-E_{l})/\hbar $.

By applying the Zwanzig's projection method~\cite{zw} on the system
of coupled equations~(\ref{rnm}), we obtain an equation of motion for
the diagonal part of the density matrix $\rho _{nn}$ \cite{stoch}
\begin{equation}
\frac{\partial \rho _{nn}(t)}{\partial t}=\int_{0}^{t}ds\sum_{m}\left\{
W_{nm}(t)W_{mn}(s)\left[ \rho _{mm}(s)-\rho _{nn}(s)\right] +c.c.\right\} ,
\label{rnn}
\end{equation}%
which determines the intrinsic excitation of the nucleus $E^{\ast }$ at time
$t$ and which should be supplemented by the normalization condition,
\begin{equation}
\sum_n \rho_{nn}=1.
\label{rnn1}
\end{equation}
$c.c.$ in Eq.~(\ref{rnn}) stands for the complex conjugation.

Using Eqs.~(\ref{wnl}) and~(\ref%
{movbas}), functions $W_{nm}$ in Eq.~(\ref{rnn}) may be written as
\begin{equation}
W_{nm}(t)=i\dot{q}(t)\exp \left( -i\int_{0}^{t}\omega _{nm}dt^{\prime
}\right) \bigg[\frac{\langle \Psi _{n}|\partial \hat{H}/\partial q|\Psi
_{m}\rangle }{E_{n}-E_{m}}\bigg](q[t]).  \label{wnm}
\end{equation}%
Substituting these expressions into into Eq.~(\ref{rnn}) and
assuming that the matrix elements
$\langle \Psi _{n}|\partial \hat{H}/\partial q|\Psi _{m}\rangle$ and energy
distances $E_n-E_m$ rapidly fluctuate with time, one has
\begin{equation}
\frac{\partial \rho _{nn}(t)}{\partial t}=2\dot{q}(t)\sum_{m}\left[ \frac{%
|\langle \Psi _{n}|\partial \hat{H}/\partial q|\Psi _{m}\rangle |^{2}}{%
(E_{n}-E_{m})^{2}}\right] \int_{0}^{t}ds\dot{q}(s)
cos\bigg([E_n-E_m]/\hbar[t-s]\bigg)[\rho _{mm}(s)-\rho
_{nn}(s)].  \label{ww}
\end{equation}

At high excitation energies, the nuclear spectrum is very complex
and can be described by a random matrix theory~\cite{bm}. In
the random matrix theory approach we ensemble average randomly
distributed energy distances $E_n-E_m$ and $E_{m}$ and squared
off-diagonal matrix elements $|h|_{nm}^{2}\equiv |\langle \Psi
_{n}|\partial \hat{H}/\partial q|\Psi _{m}\rangle |^{2}$,
\begin{eqnarray}
\frac{\partial \bar{\rho}(E_{n},t)}{\partial t} &=&2\dot{q}%
(t)\int_{0}^{t}ds\ \dot{q}(s)\int_{E_{gs}}^{+\infty }dE_{m}\
R(|E_{n}-E_{m}|)\Omega (E_{m})  \nonumber \\
&&\times \frac{\cos \{(E_{n}-E_{m})/\hbar (t-s)\}}{(E_{n}-E_{m})^{2}}%
\overline{|h|^{2}}(E_{n},E_{m},q[t])[\bar{\rho}(E_{m},s)-\bar{\rho}%
(E_{n},s)],  \label{rdis}
\end{eqnarray}%
where "bar" above a quantity means the corresponding ensemble averaged value
and $\Omega $ is the average nuclear level--density. Here, it was assumed
that the ensemble averaging over the energy distances and squared
off-diagonal matrix elements can be performed independently. In Eq.~(\ref%
{rdis}), $R(|E_n-E_m|)$ is the two--level correlation function which
is the probability density to find the state $m$ with energy $E_m$ in
interval $[E_m,E+dE_m]$ at the average distance $|E_n-E_m|$
from the given state $n$ with energy $E_n$.

In the nuclear case, the many--body Hamiltonian $\hat{H}$ obeys time--reversal
symmetry implying the usage of Gaussian Orthogonal Ensemble (GOE) to
model the nuclear spectrum. For a general mesoscopic system~\cite{ge}, $\hat{H}$
may not have a time--reversal symmetry and one has to use Gaussian
Unitary Ensemble (GUE) of many--body levels. Correspondingly,

\bs

(i) For the GOE statistics~\cite{pm}
\begin{equation}
R_{GOE}(x)=1-\left( \frac{\mathrm{sin}(\pi x)}{\pi x}\right) ^{2}+\left(
\int_{0}^{1}dy\frac{\mathrm{sin}(\pi xy)}{y}-\frac{\pi }{2}\right) \left(
\frac{\mathrm{cos}(\pi x)}{\pi x}-\frac{\mathrm{sin}(\pi x)}{(\pi x)^{2}}%
\right) ,  \label{RGOE}
\end{equation}

\bs

(ii) while in the GUE case
\begin{equation}
R_{GUE}(x)=1-\left( \frac{\mathrm{sin}(\pi x)}{\pi x}\right) ^{2},
\label{RGUE}
\end{equation}%
where $x\equiv |E_{n}-E_{m}|\ \Omega (E_{n})$. The behaviour of the
two--level correlation function $R(x)$ with the normalized level
spacing $x$ for the different statistical ensembles~(\ref{RGOE}) and
(\ref{RGUE}) is shown in Fig.~1. The main difference between the
GOE and GUE cases, seen in Fig.~1, is the behaviour of $R(x)$ at
small energy spacings $x$. For the GOE statistics one has linear
repulsion between levels, $R_{GOE}\sim x$, while the GUE statistics
implies quadratic level repulsion, $R_{GUE}\sim x^{2}$. On the other
hand, $R_{GOE}$ and $R_{GUE}$ are similar at moderate spacings $x$,
when the spectral correlations between levels consistently disappear.

The next ingredient of the statistical averaging procedure is the ensemble
averaged squared matrix elements (EASME) $\overline{|h|^{2}}$. It is rather
clear that at high excitation energies the transition matrix elements
between the complex many--body states should drop out with the energy
distance between them. In order to characterize the $\overline{|h|^{2}}%
(|E_n-E_m|)$--distribution, we introduce the strength of the
distribution $\sigma ^{2}$ and its width $\Gamma$. To clarify the
physical meaning of the quantities $\sigma^2$ and $\Gamma$, one
may use the random matrix approach of Ref.~\cite{ra}, where the
nuclear many--body states are constructed on unperturbed basis
states which are linearly coupled to external time--dependent
classical variable $q(t)$, and complexity is achieved by adding
two--body interaction between them. In this approach $\sigma^2$
is the variance of the slopes, $\partial E_n/\partial q$, of the
many--body energy levels.
The strength of the two--body interaction, introduced to model
the effect of residual interaction between nucleons, defines
the spreading width $\Gamma_{\mu}$ of the squared off--diagonal matrix
elements $|\langle n|\partial\hat{H}/\partial q|m \rangle|^2$,
for example, via Fermi's Golden Rule.

Thus, we present the ensemble averaged squared matrix elements
(EASME) $\overline{|h|^{2}}$ in the following form
\begin{equation}
\overline{|h|^{2}}=\frac{\sigma ^{2}}{\sqrt{\Omega (E_{n})\Omega (E_{m})}%
\Gamma }f(|E_{n}-E_{m}|/\Gamma ),  \label{h2f}
\end{equation}%
where $f$ is a shape of the EASME's distribution with the natural
boundary conditions, $f(0)=const$ and
$f(\infty )=0$.

Going from the discrete energy levels $E_{n}$ and $E_{m}$ to continuous
energy variables,
\begin{equation}
E\equiv E_{n},~~~e\equiv E_{m}-E_{n},  \label{Ee}
\end{equation}%
and substituting the expression~(\ref{h2f}) into Eq.~(\ref{rdis}), we obtain
\begin{eqnarray}
\frac{\partial \bar{\rho}(E,t)}{\partial t}=
\frac{\sigma^2}{\sqrt{\Omega (E)}\Gamma}
\dot{q}(t)\int_{0}^{t}ds\ \dot{q}(s)\times
\int_{-\infty}^{+\infty }\frac{f(|e|/\Gamma )}{e^{2}}
\nonumber \\
R(\Omega (E)|e|)cos(e/\hbar (t-s))\sqrt{\Omega (E-e)}[\bar{\rho}(E-e,s)-
\bar{\rho}(E,s)]de.  \label{rcont}
\end{eqnarray}%
In the last equation, the integration limits over the energy spacing $e$
were extended to infinities since the time changes of
the occupancy $\bar{\rho}(E,t)$ of the given nuclear state with the energy $%
E $ are mainly due to the direct interlevel transitions from the
close--lying states located at the distances $|e|<<E$. The same
assumptions enable us to truncate expansion to $e^3$--order
terms,
\begin{eqnarray}
\sqrt{\Omega (E-e)}[\bar{\rho}(E-e,s)-\bar{\rho}(E,s)] &=&-\sqrt{\Omega (E)}%
\frac{\partial \bar{\rho}(E,s)}{\partial E}e  \nonumber \\
+\frac{1}{2\sqrt{\Omega (E)}}\frac{d\Omega (E)}{dE}\frac{\partial \bar{\rho%
}(E,s)}{\partial E}e^{2}+\frac{\sqrt{\Omega (E)}}{2}\frac{\partial ^{2}\bar{%
\rho}(E,s)}{\partial E^{2}}e^{2}+(...)e^3 + O(e^4)
\label{expan}
\end{eqnarray}
Substituting the expansion~(\ref{expan}) into Eq.~(\ref{rcont}),
the odd-$e$ terms drop out and dynamical equation for the
occupancy $\bar{\rho}(E,t)$ of the nuclear state with the energy
$E$ becomes
\begin{equation}
\Omega(E)\frac{\partial \bar{\rho}(E,t)}{\partial t}\approx \sigma^{2}
\dot{q}(t)\int_{0}^{t}ds K(t-s)\dot{q}(s)\frac{\partial }{%
\partial E}\bigg[\Omega (E)\frac{\partial \bar{\rho}(E,s)}{\partial E}\bigg],
\label{rnonm}
\end{equation}%
where retardation of the $\bar{\rho}(E,t)$--dynamics is defined by
a memory kernel $K(t-s)$ which is defined by the Fourier
transform of the product of the EASME's energy distribution
$f$ and the two--level correlation function $R$,
\begin{equation}
K(t-s)=\frac{1}{\Gamma}Re\bigg(\int_{\infty}^{+\infty}
f(|e|/\Gamma)R(|e|\Omega(E))exp(\frac{ie[t-s]}{\hbar})de\bigg).
\label{K}
\end{equation}

Non--Markovian equation~(\ref{rnonm}) describes the process of energy
diffusion in the space of highly excited many--body states. The memory
effects in the intrinsic energy diffusion is defined by the counterplay
between a time spread of the memory kernel~(\ref{K}), $\tau\sim \hbar/\Gamma$,
and a typical collective time $\tau_{coll}$ (a duration of the
physical process). Depending on the width $\Gamma$
of the EASME's energy distribution~(\ref{h2f}), we distinguish different
dynamical regimes of the intrinsic energy diffusion~(\ref{rnonm}):

\bs

(i) $\hbar/\Gamma<<\tau_{coll}$. In this case, $K(t-s)$ is
sharply peaked around $t=s$, and one can integrate by parts the
right--hand side of Eq.~(\ref{rnonm}) and keep only leading order
term in a small parameter $\hbar/\Gamma$. Thus, we obtain a
Markovian limit of the intrinsic dynamics~(\ref{rnonm}):
\begin{equation}
\Omega(E)\frac{\partial \bar{\rho}(E,t)}{\partial t}\approx \frac{\hbar\sigma
^{2}f(0)}{\Gamma}\dot{q}^{2}(t)\frac{\partial }{\partial E}
\bigg[\Omega (E)\frac{\partial \bar{\rho}(E,t)}{\partial E}\bigg].
\label{r8}
\end{equation}%
Here, the intrinsic energy diffusion is determined by the diffusion
coefficient $D_{E}=\hbar\sigma^2 f(0)\dot{q}^{2}/\Gamma$ which
grows with the square of the collective velocity $\dot{q}$ and drops
out with the increase of the width $\Gamma$. The latter
feature of the quantum mechanical energy diffusion can be understood
as follows. The width $\Gamma$ of the EASME's energy distribution
defines an effective number of states $N\sim\Gamma\Omega(E)$ coupled
by the transition operator $\partial\hat{H}/\partial q$ at the given
excitation $E$. The initially occupied many--body state with
energy $E$ will spread out over $N$ neighboring
states, resulting in a gradual equilibration of the quantum
mechanical intrinsic subsystem. The larger $\Gamma$, the
closer the intrinsic subsystem to the equilibrium and therefore,
the weaker the energy diffusion.

\bs

(ii) $\hbar/\Gamma>>\tau_{coll}$. Now we can put approximately
$K(t-s)\approx K(0)$ for the memory kernel in Eq.~(\ref{rnonm})
and get
\begin{equation}
\Omega(E)\frac{\partial \bar{\rho}(E,t)}{\partial t}\approx
\sigma^{2}K(0)\dot{q}(t)\Delta q(t)\frac{\partial }{\partial E}
\bigg[\Omega (E)\frac{\partial \bar{\rho}(E,t)}{\partial E}\bigg],
\label{r0}
\end{equation}%
where $\Delta q(t)=q(t)-q(t=0)$ is the change of collective
deformation of the nucleus. Here the diffusion coefficient 
$D_E = \sigma^2 K(0) \dot{q} \Delta q$ is linearly proportional
to the collective velocity $\dot{q}$ and does not depend on
the width $\Gamma$.

\bigskip

(iii) $\hbar/\Gamma \sim \tau_{coll}$. In the intermediate
case, the memory effects in the intrinsic energy diffusion~(\ref{rnonm})
will be of maximal size.

\bs

It is natural to address a question of the effect of level statistics
(\ref{RGOE})--(\ref{RGUE}) on the intrinsic energy diffusion. We believe
that the energy diffusion will differ significantly for the statistical
ensembles of levels~(\ref{RGOE})--(\ref{RGUE}) only at quite small
values of the width $\Gamma$, $\Gamma \leq 1/\Omega$, i. e., when the
features of
the nuclear spectrum at small spacings between levels show up; see Fig.~1.
On the other hand, at quite large widths $\Gamma>>1/\Omega$ the spectral
statistics effect is of a minor role as far as the statistical ensembles
of levels~(\ref{RGOE})--(\ref{RGUE}) show the universal behavior at large
level spacings. The latter regime is realized for the highly excited nuclei
provided that the width $\Gamma $ of the EASME's distribution~(\ref{h2f})
may lie in a quite wide energy interval $\Gamma \sim (10^{0}\div 10^{6})~eV$.

We may illustrate quantitatively our general discussion of the
intrinsic energy diffusion by calculating the memory kernel~(\ref{K}) for
a Lorentzian shape $f$ of the EASME's energy distribution,
\begin{equation}
f(|e|/\Gamma )=\frac{1/\pi }{(e/\Gamma )^{2}+1}.  \label{loren}
\end{equation}
To estimate the spectral statistics effect, we evaluated the memory kernel
$K(t-s)$ at $s=t$ for the different levels ensembles~(\ref{RGOE})--(\ref{RGUE}).
The corresponding results for $K(0)$ as a function of the reduced width $\Gamma
\Omega (E)$ are shown in Fig.~2. As was discussed above, the level
statistics play a role only for quite small parameters $\Gamma \Omega (E)$
and the effect from the spectral statistics on the intrinsic energy
diffusion~(\ref{rnonm}) disappears at large widths $\Gamma \Omega (E)$.

For $\Gamma\Omega>>1$, one can find analytically
the memory kernel~(\ref{K})
\begin{equation}
K(t-s)=exp\left( -\frac{|t-s|}{\hbar /\Gamma }\right) ,  \label{exp}
\end{equation}%
leading to the following non--Markovian equation of motion for the occupancy
$\bar{\rho}(E,t)$ of the given nuclear state $E$,
\begin{equation}
\Omega(E)\frac{\partial \bar{\rho}(E,t)}{\partial t}=\frac{\sigma ^{2}}{\Gamma}
\dot{q}(t)\int_{0}^{t}exp\left( -\frac{|t-s|}{\hbar /\Gamma }\right)
\dot{q}(s)\frac{\partial }{\partial E}\bigg[\Omega (E)\frac{\partial \bar{%
\rho}(E,s)}{\partial E}\bigg]ds.  \label{rnonml}
\end{equation}

\section{Intrinsic diffusion and collective dissipation}

\label{idcd}

Now we are able to obtain a dynamical equaion for the intrinsic excitation
energy of the nucleus $E^{*}(t)$,
\begin{equation}
E^{\ast }(t)=\sum_{n}E_{n}\bar{\rho}_{nn}(t)=\int_{0}^{+\infty }dE\
\Omega (E)E\bar{\rho}(E,t),  \label{E*}
\end{equation}
which enters the equation of motion~(\ref{colem}) for the
classical collective variable $q(t)$. By using Eq.~(\ref{rnonm}),
one gets after partial integration
\begin{equation}
\frac{dE^{*}}{dt}=\frac{\sigma ^{2}}{\Gamma}
\dot{q}(t)\int_{0}^{t}exp\left(-\frac{|t-s|}{\hbar /\Gamma }\right)
\dot{q}(s)\int_0^{+\infty}\frac{d\Omega(E)}{dE}\bar{\rho}(E,s)dEds
\label{rnonml}
\end{equation}

We stress immediately that the collective motion is undamped for the
constant nuclear level--density, $\Omega (E)=const$. In that case the
intrinsic subsystem
is not excited during the collective motion, $E^{*}(t)=E^{*}(t=0)$ and
therefore, due to the energy conservation condition~(\ref{Etot}), the
collective energy is constant in time. This means that the growth of
the average nuclear level--density $\Omega$ with energy is the necessary
condition for the collective dissipation. In the sequel, we will use the
constant--temperature level--density,
\begin{equation}
\Omega (E)=c\cdot exp(E/T),  \label{T}
\end{equation}%
where $T$ is the temperature of the nucleus, and which leads us to
non--Markovian collective dynamics,
\begin{equation}
B(q)\ddot{q}(t)=-\frac{1}{2}\frac{\partial B(q)}{\partial q}\dot{q}^{2}(t)-%
\frac{\partial E_{pot}(q)}{\partial q}-\frac{\sigma ^{2}}{T}\int_{0}^{t}
exp\left( -\frac{|t-s|}{\hbar /\Gamma }\right) \dot{q}(s)ds.
\label{qnonm}
\end{equation}
We see from Eq.~(\ref{qnonm}) that the non--Markovian character of the
intrinsic nuclear dynamics~(\ref{rnonm}) gives rise to the presence of
memory effects in the macroscopic collective motion. Correspondingly,
the Markovian limits of the intrinsic energy diffusion~(\ref{r8}) and
(\ref{r0}) would correspond to the Markovian collective motion. Indeed,
for the quite broad energy distributions of the EASME~(\ref{h2f}),
$\hbar/\Gamma<<\tau_{coll}$, we obtain
\begin{equation}
B(q)\ddot{q}(t)=-\frac{1}{2}\frac{\partial B(q)}{\partial q}\dot{q}^{2}(t)-%
\frac{\partial E_{pot}(q)}{\partial q}-\frac{\hbar\sigma ^{2}}{\Gamma T }%
\dot{q}(t).  \label{q8}
\end{equation}%
Here an ordinary friction force with the friction coefficient $\hbar \sigma
^{2}/(\Gamma T)$ appears as a result of the Markovian intrinsic energy
diffusion~(\ref{r8}).

In the opposite case of the intrinsic dynamics~(\ref{r0}), when the EASME's
distribution is strongly peaked, $\hbar/\Gamma>>\tau_{coll}$, we obtain
a friction--less limit of the collective motion,
\begin{equation}
B(q)\ddot{q}(t)=-\frac{1}{2}\frac{\partial B(q)}{\partial q}\dot{q}^{2}(t)-
\frac{\partial E_{pot}(q)}{\partial q}-\frac{\sigma^2}{T}(q(t)-q_0),  \label{q0}
\end{equation}
when the retarded force in the right--hand side of Eq.~(\ref{qnonm}) is
reduced to a pure conservative force $\sigma^2(q-q_0)/T$.

\section{Nuclear fission calculations}

\label{nfc}

Even within a very simple one--dimension model for the collective dynamics (%
\ref{qnonm}), we may calculate quantities which can be estimated
from experimental observables. Let us consider a symmetric
fission of the highly excited $^{236}U$. The classical collective
variable $q(t)$ can be chosen as the elongation of axial symmetric
nuclear shape measured in units of the radius $R_{0}=r_{0}A^{1/3}$
of the nucleus. The collective potential energy from saddle
point to scission $E_{pot}(q)$, shown in Fig.~3, is approximated
by an inverted parabolic potential~\cite{kramers,nix},
\begin{equation}
E_{pot}(q)=8~MeV-\frac{1}{2}\hbar \omega _{f}B(q_{0})(q-q_{0})^{2},
\label{Egs}
\end{equation}%
where $\hbar \omega _{f}=1.16~MeV$, $q_{0}$ is the initial 
(saddle-point) deformation of the nucleus, $q_{0}=q(t=0)=1.6$,
and $B(q)$ is the collective mass coefficient derived for the
incompressible and irrotational nuclear fluid,
\begin{equation}
B(q)=\frac{1}{5}AmR_{0}^{2}(1+\frac{1}{2q^{3}}),  \label{Bq}
\end{equation}%
with the nucleonic mass $m$. The scission point $q_{sc}$ can be obtained
from the following condition~\cite{nix}
\begin{equation}
E_{pot}(q_{0})-E_{pot}(q_{sc})=20~MeV.  \label{qsc}
\end{equation}%
The initial collective kinetic energy is taking to be equal to $1~MeV$.

Characterizing the intrinsic nuclear motion, we adopt the initial
temperature of the nucleus $T=2~MeV$ and estimate the strength $\sigma ^{2}$
of the EASME's distribution within the Nilsson model for single--particle
nuclear states in an anisotropic harmonic oscillator potential, see Ref.~%
\cite{willets}:
\begin{equation}
\sigma ^{2}=\frac{3m^{2}\omega _{0}^{3}AR_{0}^{4}}{560\pi \hbar },
\label{s2}
\end{equation}%
with $\hbar \omega_0 =41/A^{1/3}~MeV$.

Using Eq.~(\ref{qnonm}), we calculated numerically from Eq.~(\ref{qnonm}%
) the time, $t_{sc}$, of the nuclear descent from the top of fission barrier
$q_{0}$ to the scission point $q_{sc}$ (\ref{qsc}). The corresponding
results for the saddle--to--scission time $t_{sc}$ are plotted in Fig.~4 as
a function of the width $\Gamma $ of the Lorentzian distribution of the
EASME~(\ref{h2f}). As can be seen from Fig. 4, the time for the nuclear
descent $t_{sc}$ decreases with the increase of the width $\Gamma $ of the
EASME's distribution~(\ref{h2f}). In order to explain such kind of behavior
of $t_{sc}$, we represent the retarded force in the right--hand side of Eq.~(%
\ref{qnonm}) as a sum,
\begin{equation}
-\frac{\sigma^{2}}{T}\int_{0}^{t}exp\left( -\frac{|t-s|}{\hbar
/\Gamma }\right) \dot{q}(s)ds=-\gamma (t,\hbar /\Gamma )\dot{q}(t)-\tilde{C}%
(t,\hbar /\Gamma )(q(t)-q_{0}),  \label{retar}
\end{equation}%
where $\gamma $ and $\tilde{C}$ are the time--dependent friction and
stiffness coefficients, respectively. The separation~(\ref{retar}) of the
retarded force is general in the sense that it always contains the
time--irreversible (the friction part) and time--reversible (the
conservative part) contributions. In fact, the memory effects in the
collective dynamics~(\ref{qnonm}) give rise to the friction, $\gamma (t)\dot{%
q}(t)$, and lead to the renormalization of the stiffness of the nuclear
many--body system,
\begin{equation}
C=-B(q_{0})(\hbar \omega _{f})^{2}+\tilde{C}(t,\hbar /\Gamma ),  \label{cc}
\end{equation}%
see Eqs.~(\ref{qnonm}) and~(\ref{retar}). It is important that $\tilde{C}$
is always positive resulting in the additional hinders of the nuclear
descent from the fission barrier, see Ref.~\cite{krs}. The relative sizes of
the friction and the dynamic conservative forces in~(\ref{retar}) are
defined by the time--spread of the exponential kernel, $\hbar /\Gamma $. If
the dynamic stiffness $\tilde{C}$ is expected to increase monotonically with
$\hbar /\Gamma $, the friction coefficient $\gamma $ is a non--monotonic
function of the memory time $\hbar /\Gamma $. At the limit of relatively
small values of $\hbar /\Gamma $ (the large--widths limit which we consider
here), both the friction and the dynamic conservative
contributions drop out with the memory time explaining the decay of the
saddle--to--scission time $t_{sc}$ with the width $\Gamma $ of the EASME's
distribution.

By using our previous estimations of the saddle--to--scission time done in
Ref.~\cite{krs} for the same one--parametric nuclear shape parameterization~(%
\ref{Egs})--(\ref{qsc}), $t_{sc}\sim (6\div 12)\cdot 10^{-21}s$, we can
conclude from Fig.~4 that the width $\Gamma $ of the Lorentzian distribution
of the EASME lies in the interval $10~MeV\leq \Gamma \leq 20~MeV$.

We also calculated the dependence of collective kinetic energy at the
scission point $E_{ps}$ on the width $\Gamma $, see Fig.5. As far as the
nuclear descent gets faster with the width of the EASME's distribution, the
collective energy of the nucleus at the scission point will increase with $%
\Gamma $. The estimated interval for the width, $10~MeV\leq \Gamma \leq
20~MeV$, obtained from our saddle--to--scission calculations (see Fig.~4),
gives realistic values of the pre--scission kinetic energy $1~MeV\leq
E_{ps}\leq 3~MeV$ \cite{krs}.

\section{Summary}

\label{summ}

In attempt to describe selfconsistently the nuclear many--body dynamics
undergoing along macroscopic collective path and intrinsic excitations, we
have applied the cranking approach (\ref{Etot})--(\ref{colem}) to the
nucleus. We have introduced a single time--dependent classical variable $%
q(t) $ to characterize the slow collective nuclear motion, while the fast
intrinsic modes of the motion have been treated quantum--mechanically within
the Liouville equation~(\ref{rho}) for the nuclear density matrix. Applying
the Zwanzig's projection method~\cite{zw}, the intrinsic nuclear dynamics
has been reduced to the equation of motion~(\ref{rdis}) for the occupancies
of nuclear states. To obtain Eq.~(\ref{rdis}), we have averaged the
intrinsic dynamics over the randomly distributed level spacings $e$ and
squared matrix elements $|h|^{2}$ of the transition operator $\partial \hat{H%
}/\partial q$, where $\hat{H}$ is the nuclear many--body Hamiltonian. The
distribution of the ensemble averaged matrix elements (EASME) $\overline{%
|h|^{2}}$ has been taken in the general form~(\ref{h2f}), where decay of $%
\overline{|h|^{2}}$ with the energy distance between states $e$ has been
characterized by the width $\Gamma $ and we have assumed that $\overline{%
|h|^{2}}$ drops out strongly with excitation energy through the average
nuclear level--density $\Omega $.

In the limit of high excitations of the nucleus, we have obtained the
non--Markovian diffusion equation~(\ref{rnonm}) for the time evolution of
the intrinsic nuclear occupancies. The time features of the intrinsic energy
diffusion~(\ref{rnonm}) are defined by the relation between the time scale
$\hbar/\Gamma$ and the characteristic time scale of the collective motion
$\tau_{coll}$. We have found that at fairly broad energy distributions
of the EASME~(\ref{h2f}), i. e., when $\hbar/\Gamma<<\tau_{coll}$,
Markovian regime~(\ref{r8}) of the intrinsic energy diffusion is
observed with the diffusion coefficient quadratically depending on the
collective velocity $\dot{q}$ and inversely proportional to the width $%
\Gamma $. In the opposite case of quite small widths $\Gamma$,
$\hbar/\Gamma>>\tau_{coll}$, we also found the normal (Markovian)
regime of the intrinsic energy diffusion but with the diffusion
coefficient linearly proportional to the collective velocity $\dot{q}$
and not depending on the width $\Gamma$.

We have investigated how the level spacing statistics can influence the
intrinsic energy diffusion~(\ref{rnonm}). Only in the case of quite small
widths of the EASME's distribution, $\Gamma \Omega \leq 1$, the significant
difference of the intrinsic dynamics for the Gaussian orthogonal (GOE)~(\ref%
{RGOE}) and Gaussian unitary (GUE)~(\ref{RGUE}) ensembles of levels is
expected. Such a difference would disappear as far as the product $\Gamma
\Omega $ becomes larger and larger. We may explain that by the fact that the
transitions between the nuclear states may be sensitive to the level
statistics only when the coupling between states is of order of the average
level spacing, $\Gamma \sim 1/\Omega $, i.e. when the different
small--spacing behavior of the GOE and GUE statistics may shows up. At high
nuclear excitations, we have believed that the product $\Gamma \Omega >>1$
and therefore, one can neglect the role of the spectral statistics on the
intrinsic energy diffusion~(\ref{rnonm}). We have illustrated quantitatively
this feature of the intrinsic dynamics by applying the Lorentzian
distribution~(\ref{loren}) of the EASME~(\ref{h2f}), see Fig.~3.

Our next goal was to calculate the nuclear fission's
characteristics within our approach. By using the
constant--temperature level--density~(\ref{T}), we have
derived non--Markovian equation of motion~(\ref{qnonm}) for the
classical collective variable $q(t)$, where the influence of the
intrinsic quantum motion on the collective dynamics is determined
by the retarded friction force. Then the non--Markovian collective
dynamics~(\ref{qnonm}) has been applied to describe descent of the
nucleus $^{236}U$ from the top of fission barrier to the scission
point approximating the collective potential energy
on this path by the inverted parabolic potential~(\ref{Egs}) \cite%
{kramers,nix}. We have calculated the time of the nuclear descent, $t_{sc}$
(Fig.~4), and the collective kinetic energy at the scission point, $E_{ps}$
(Fig.~5), as a function of the width $\Gamma $ of the EASME's distribution.
We have found that the nuclear descent is hindered with the decrease of $%
\Gamma $ due to the ordinary friction force contribution and the additional
conservative dynamic force caused by the presence of memory effects in Eq.~(%
\ref{qnonm})~\cite{krs}. The relative size of the memory effects decreases
with the width of the EASME's distribution and, at $\Gamma \rightarrow
\infty $, we have friction--less limit of the collective motion, see Eq.~(%
\ref{q0}). From the calculations of the saddle--to-scission time and the
pre--scission kinetic energy we have estimated the value of the width $%
\Gamma $, $10~MeV\leq \Gamma \leq 20~MeV$, which is consistent with the
previous estimations of the analogous quantity done in Refs.~\cite%
{brink,zelev}.

\section{Acknowledgments}

One of us (S.V.R.) would like to thank the Department of Mathematical
Physics of the Lund Institute of Technology, University of Lund for
the kind hospitality.

\section{Figure captions}

Fig.~1: The two--level correlation function $R(x)$ vs the normalized level
spacing $x$ for the different Gaussian ensembles~of Eqs. (\ref{RGOE}) and (%
\ref{RGUE}) of energy levels.

\bigskip

Fig.~2: Dependence of the non--Markovianity of the intrinsic nuclear
dynamics~(\ref{rnonm}) on the reduced width $\Gamma \Omega (E)$ of the
Lorentzian distribution~(\ref{loren}) of the EASME. The dependence is shown
for the different spectral statistics~(\ref{RGOE}) and (\ref{RGUE}).

\bigskip

Fig.~3: Dependence of the collective potential energy $E_{pot}$ on the
nuclear shape parameter $q$ during the descent from the top of fission
barrier $q_{0}$ to the scission point $q_{sc}$ (\ref{qsc}).

\bigskip

Fig.~4: The saddle--to--scission time $t_{sc}$ of the symmetric fission of
the $^{236}U$, calculated from Eq.~(\ref{qnonm}), is shown as a function of
the width $\Gamma $ of the Lorentzian distribution of the EASME~(\ref{h2f}).

\bigskip

Fig.~5: The collective kinetic energy at the scission point $E_{ps}$ vs the
width $\Gamma $ of the Lorentzian distribution of the EASME~(\ref{h2f}).

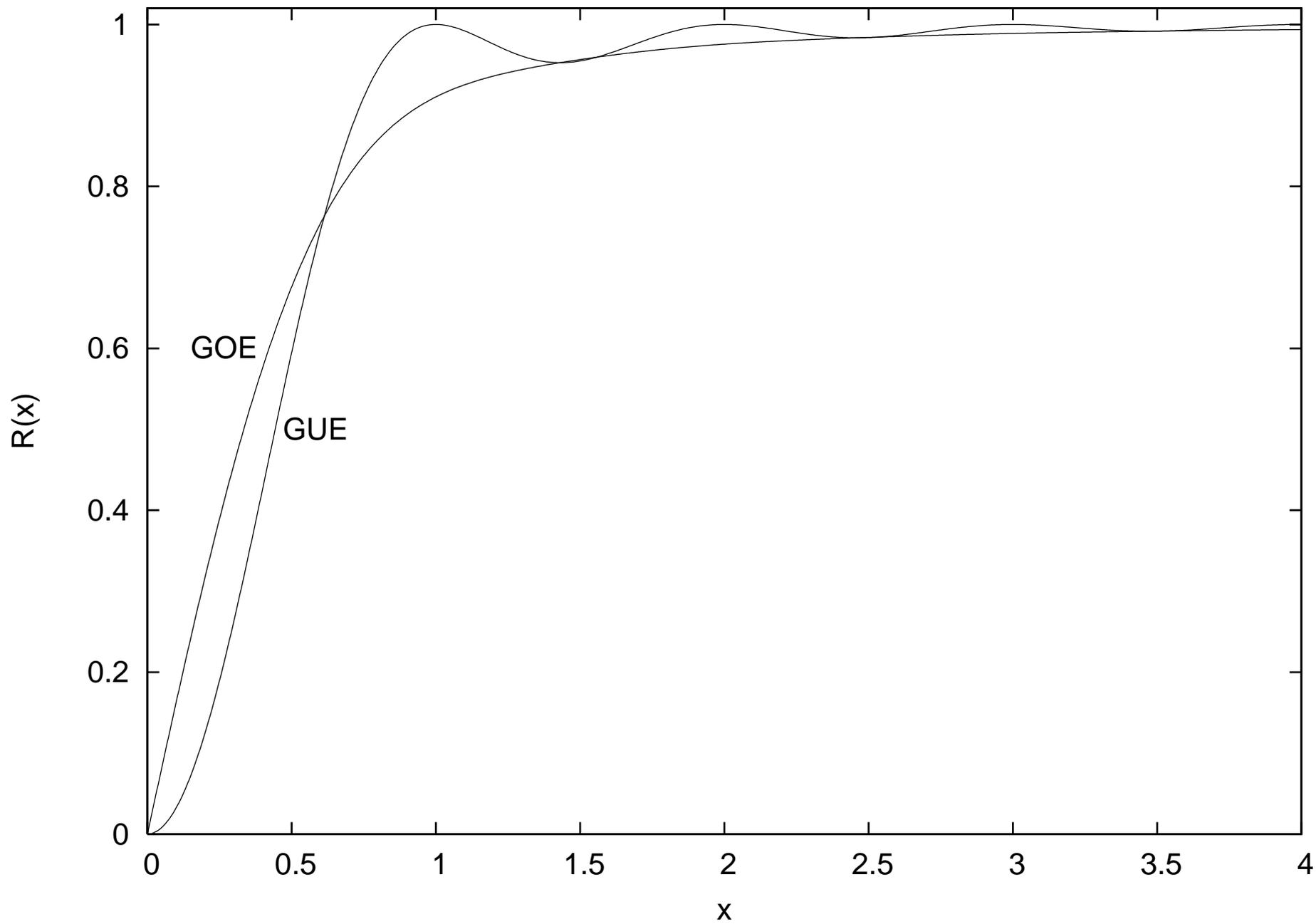
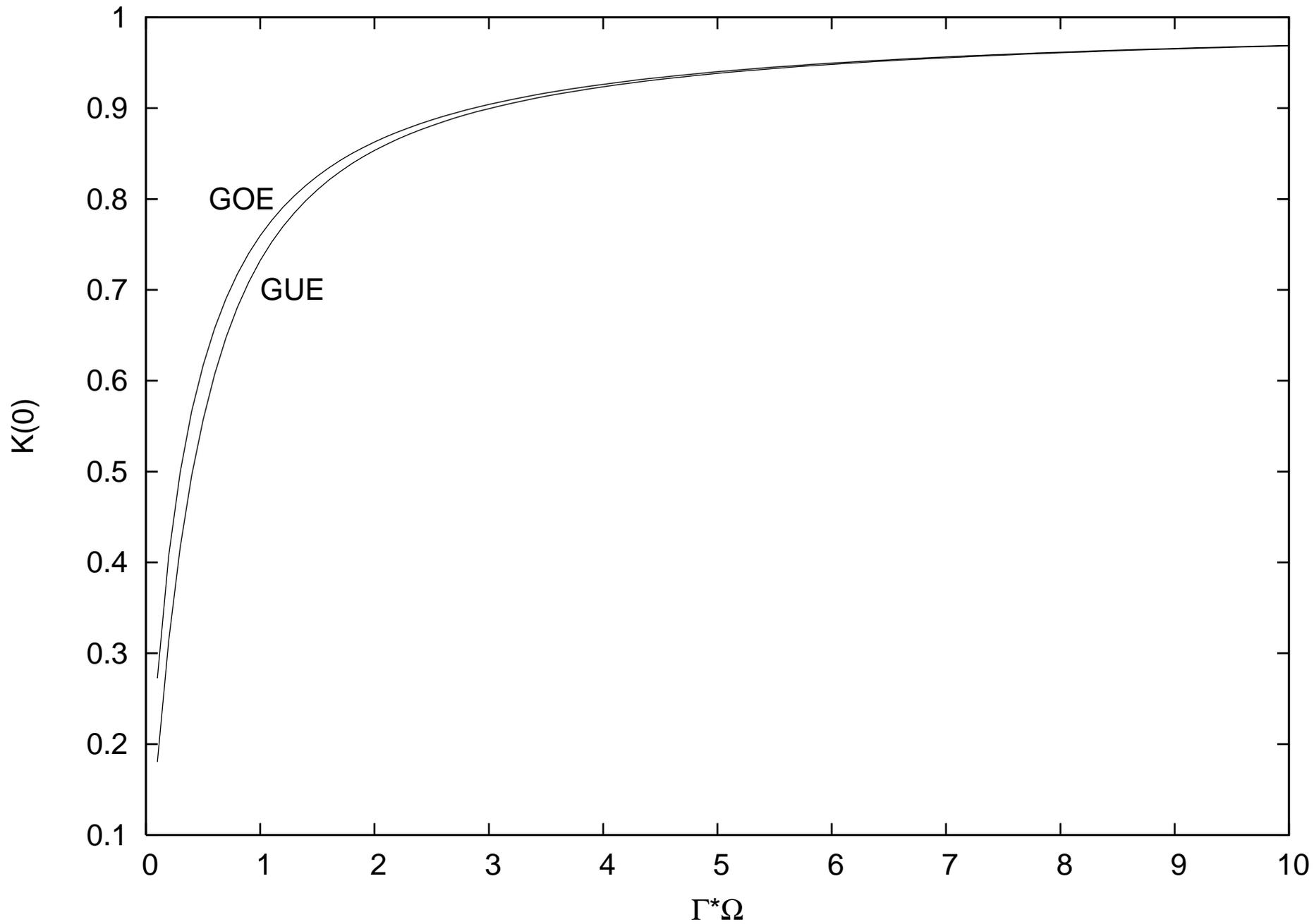
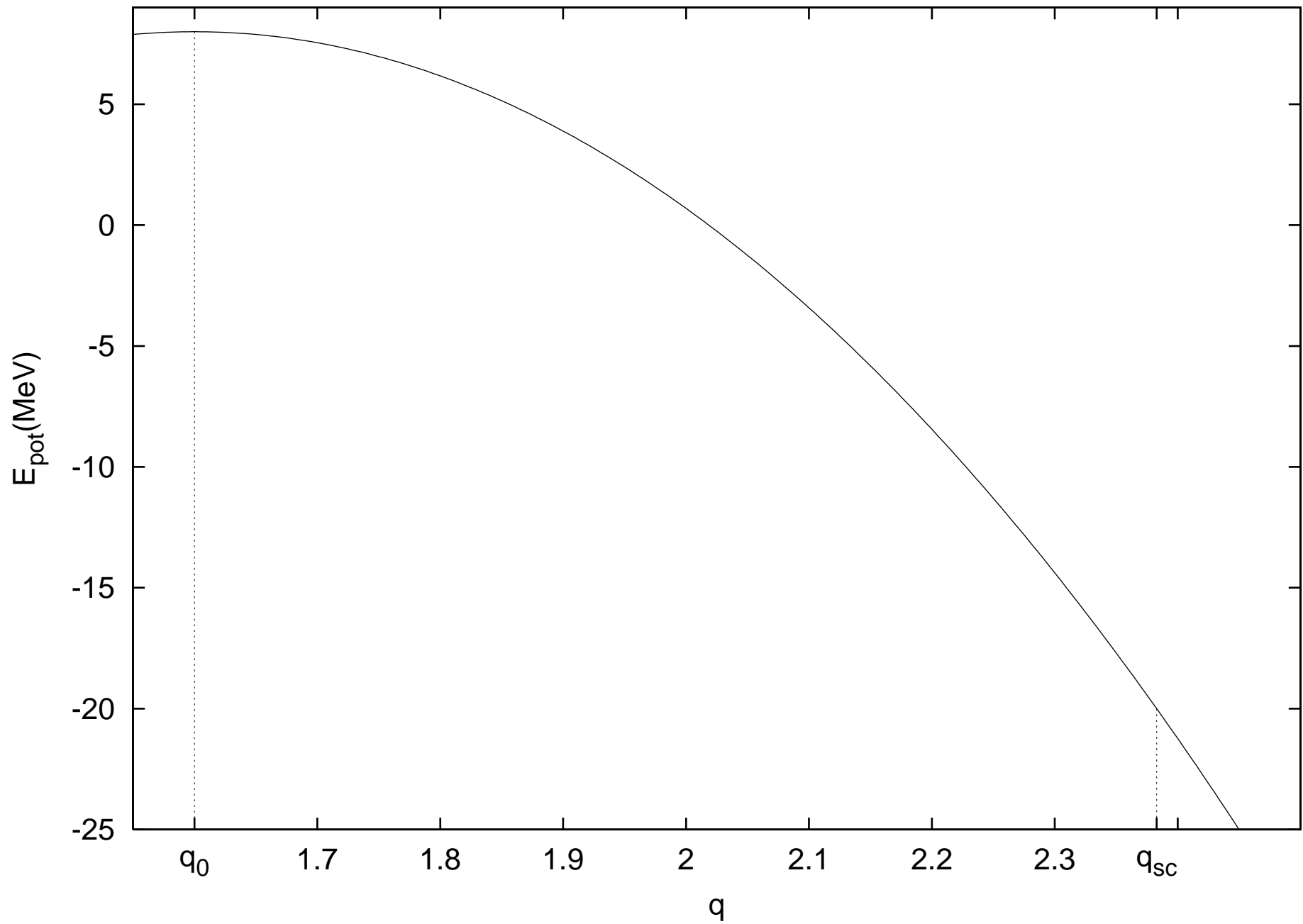
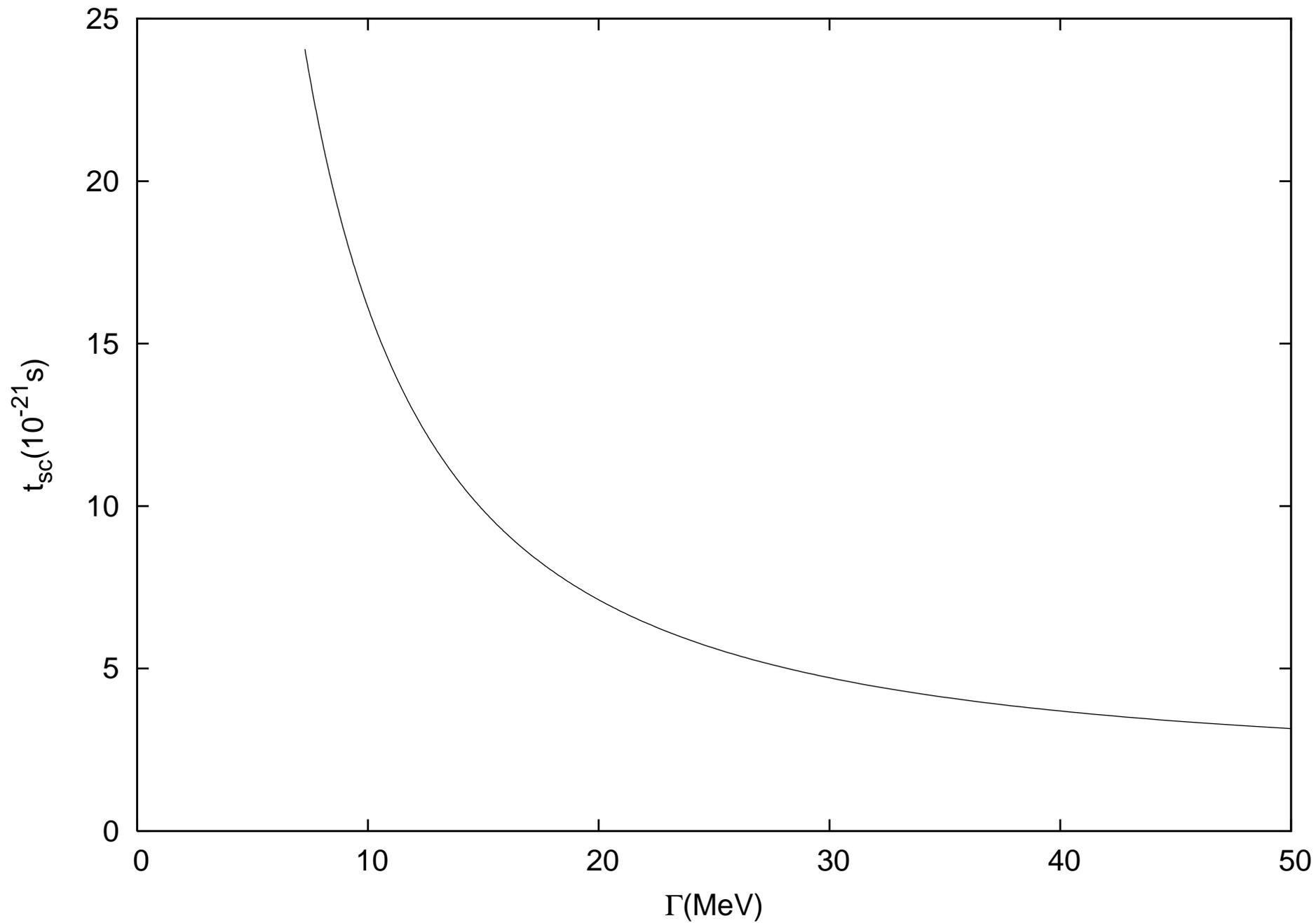
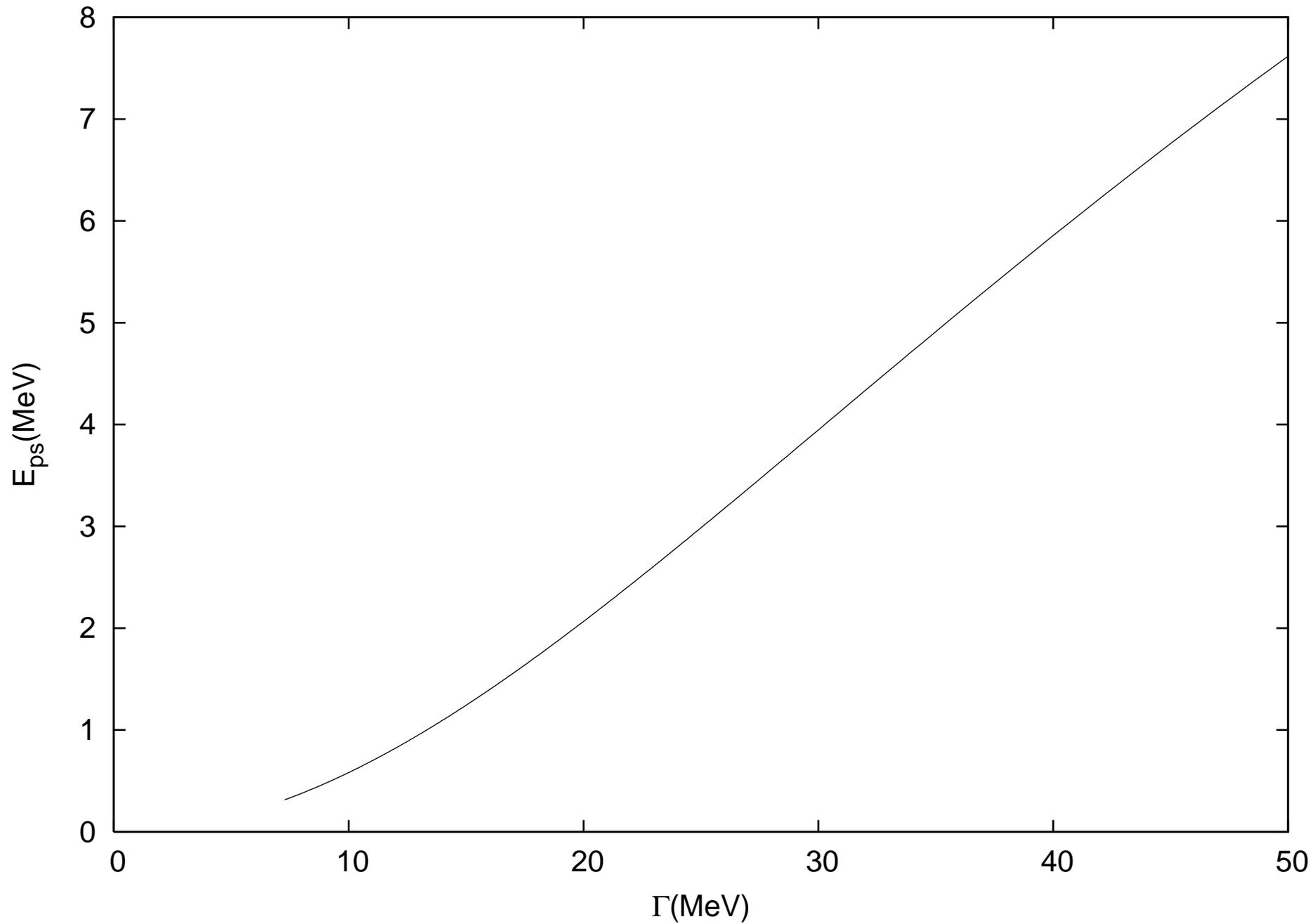
\end{document}